\documentclass[]{aa}  

\usepackage{graphicx}
\usepackage{txfonts}
\usepackage{color}
\usepackage{tikz}
\usepackage{tabu,multirow}
\usetikzlibrary{positioning}
\usepackage{amssymb}
\usepackage{amsmath}
\usepackage{pict2e}
\usepackage{multirow}
\usepackage{xcolor}
\usepackage{todonotes}
\usepackage[english]{babel}
\usepackage{marvosym}
\usepackage{natbib}
\usepackage{lipsum}
\usepackage{ulem}

\newcommand{\eye}[4]
{   \draw[rotate around={#4:(#2,#3)}] (#2,#3) -- ++(-.5*55:#1) (#2,#3) -- ++(.5*55:#1);
    \draw (#2,#3) ++(#4+55:.75*#1) arc (#4+55:#4-55:.75*#1);
    \draw[fill=gray] (#2,#3) ++(#4+55/3:.75*#1) arc (#4+180-55:#4+180+55:.28*#1);
    \draw[fill=black] (#2,#3) ++(#4+55/3:.75*#1) arc (#4+55/3:#4-55/3:.75*#1);
}

\begin{document} 
 
\title{Detection of exomoons in simulated light curves with a regularized convolutional neural network}

\author
{
Rasha Alshehhi \inst{1} 
\and
Kai Rodenbeck \inst{2,3}
\and
Laurent Gizon \inst{3,2,1}
\and
Katepalli~R.~Sreenivasan \inst{4,1}
}

\institute{
1 Center for Space Science, NYUAD Institute, New York University Abu Dhabi, Abu Dhabi, UAE
\vspace{1mm}
\\
2 Institut f\"{u}r Astrophysik,
Georg-August-Universit\"{a}t G\"{o}ttingen,
Friedrich-Hund-Platz 1, 37077 G{\"o}ttingen, Germany
\vspace{1mm}
\\
3 Max-Planck-Institut f\"ur Sonnensystemforschung, Justus-von-Liebig-Weg 3, 37077 G{\"o}ttingen, Germany
\vspace{1mm}
\\
4 Tandon School of Engineering, New York University, New York, USA
\vspace{1mm}
\\
\email{ra130@nyu.edu}
}
\date{}

\abstract
{Many moons have been detected around pla\-nets in our Solar System, but none has been detected unambiguously around any of the confirmed extrasolar planets.}
{We test the feasibility of a supervised convolutional neural network to classify  photometric transit light curves of planet-host stars and identify exomoon transits, while avoiding false positives caused by stellar variability or instrumental noise.}
{Convolutional neural networks are known to have contributed to improving the accuracy of classification tasks. The network optimization is typically performed without studying the effect of noise on the training process. Here we design and optimize a 1D convolutional neural network to classify photometric transit light curves. We regularize the network by the total variation loss in order to remove unwanted variations in the data features.}
{Using numerical experiments, we demonstrate the benefits of our network, which produces results comparable to or better than the standard network solutions. Most importantly, our network clearly outperforms a classical method used in exoplanet science to identify moon-like signals.
Thus the proposed network is a promising approach for analyzing real transit light curves in the future.}
{}

\maketitle
   
\section{Introduction}
In the past two decades, over 3,500 planets around distant stars (exoplanets) have been detected and confirmed. Most of the known exoplanets have been detected using the transit method, in which space telescopes observe a star for a period of time, generating a photometric light curve~\citep{kai_paper}. If an exoplanet passes in front of the star on its orbit around the star, light is blocked out and the brightness decreases. Planetary transits repeat every orbital period. The shape of a transit light curve, in combination with additional knowledge about the host star, is used to constrain the characteristics of the planet. If an exoplanet has a moon, it is also expected to leave a signature in the transit light curve. Detecting exomoons is difficult because a stellar variability component as well as photon noise are present~\citep{kai_paper}.

Several methods have been proposed to detect exomoons around exoplanets in light curves, using individual transits or averages over multiple transits~\citep{Kipping_2009,Heller_2014}. A planet-moon transit light curve is modeled simply as the sum of two transits; one for the planet, and a more shallow transit for the moon. To model multiple planet-moon transits, more detailed modeling is required in principle, to take the orbital motion of the moon and of the planet around their centers of mass into account. This more detailed modeling is only applied when a promising target has been identified. For previous attempts for detecting exomoons, we refer to \cite{szabo_2013} and~\cite{Hippke_2015}.{The most recent attempt has been reported by~\cite{HEK-VI-K1625}, who stacked the phase-folded transits of 284 Kepler exoplanets in search for exomoon candidates.}

\begin{figure}[t]
\centering
\begin{tikzpicture}
\draw[black,fill=yellow] (0,0) circle (1);
\node[] at (-1.6,0) {{\large star}};
\draw[gray] (-2.597,1.5) arc (-210:30:3);
\draw[->, cap=round, gray] (0,0) -- (2.1,-2.1);
\node[gray] at (1.9,-1.6) {{\large $a$}};

\draw (-1.058 ,-2.331) -- (-0.02,-4.01);
\draw[black,fill=black] (-0.681,-2.921) circle (0.1);
\draw[gray] (-0.681,-2.921) circle (1.3);
\node[] at (-0.4,-2.7) {{\large $C$}};
\draw[line width=0.1mm, black,fill=red] (-0.02,-4.01) circle (0.2);
\draw[->, cap=round, gray] (-0.681,-2.921) -- (-1.581,-3.821);
\node[gray] at (-1.5,-3.3) {{\large $a_{\rm m}$}};

\draw[black,fill=blue] (-1.058 ,-2.331) circle (0.3);
\node[black] at (-2.7 ,-2.35) {{\large planet}};
\node[black] at (0.9, -4) {{\large moon}};
\draw[gray] (-0.681,-3.1) -- (-0.681,-4);
\draw[gray, ->, cap=round] (-0.681,-3.621) arc (-90:-60:0.7);
\node[gray] at (-0.1, -3.4) {{\large $\varphi$}};
\eye{1}{0}{-6.5}{90};
\node[black] at (-1.8, -6.) {{\large observer}};

\end{tikzpicture}
\caption{Sketch of the orbital configuration for the numerical simulations of transit light curves. The planet and the moon orbit their common center of mass $C$. }
\label{fig:orbital_configuration}
\end{figure}
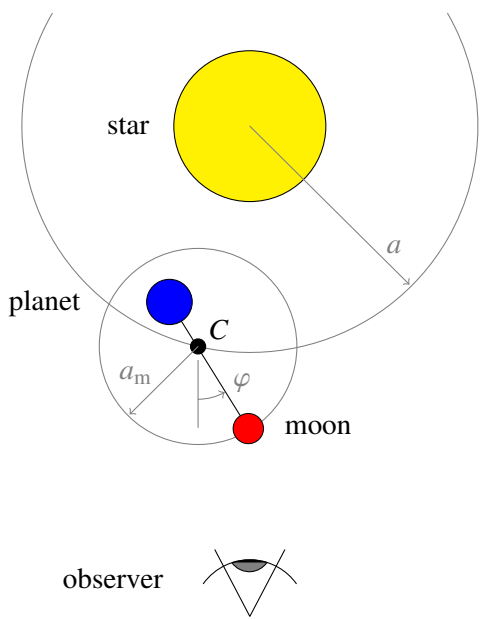

Unfortunately, no exomoon has been unambiguously detected around any exoplanet so far, although some exomoon candidates have been reported. The most promising such candidate orbits the exoplanet Kepler~1625~b. This candidate has a mass of 10\,--\,40 Earth masses (M$_{\oplus}$) and a radius of $4.90^{+0.79}_{-0.72}$ Earth radii (R$_{\oplus}$), which is  higher and larger than the masses and radiii of many confirmed exoplanets. In total, four transits of Kepler-1625~b have been observed: three by the Kepler space telescope between 2009 and 2012, and one from the Hubble space telescope in October 2017~\citep{HEK-VI-K1625,exomoon_evidence}. Kepler-1625~b orbits its star every 287 days; the exomoon orbits the planet with a period between 13 and 39 days (see Fig.~\ref{fig:orbital_configuration} for a diagram of the orbital configuration).

\begin{figure*}[t]
\centering
\includegraphics[width =\columnwidth,clip=true]{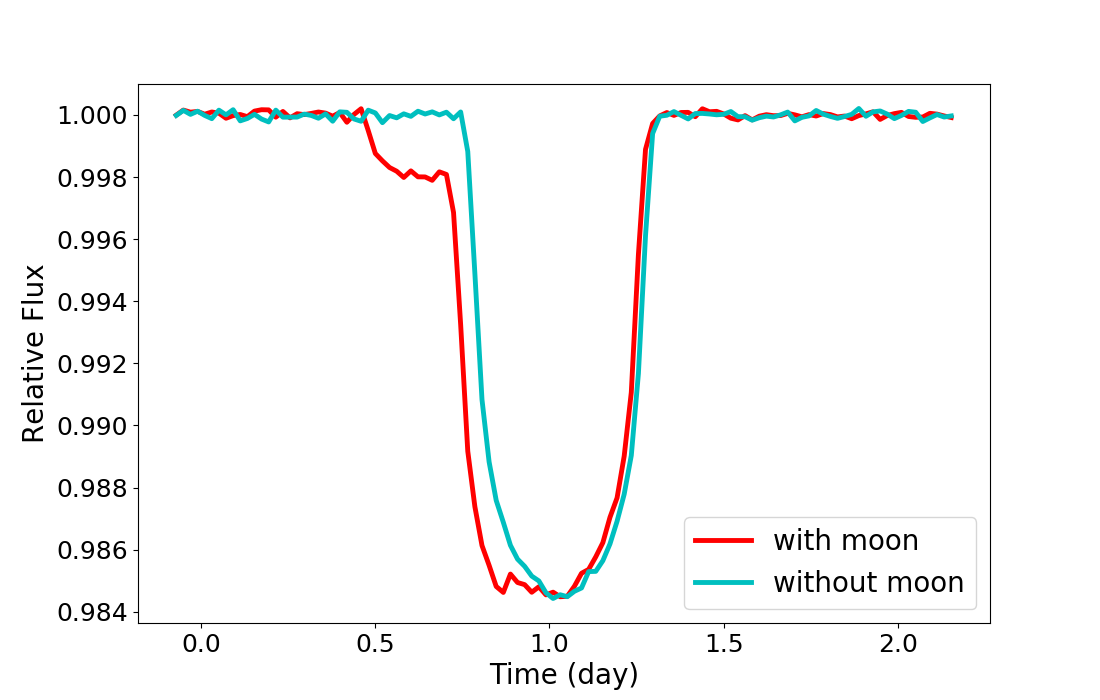}
\includegraphics[width =\columnwidth, clip=true]{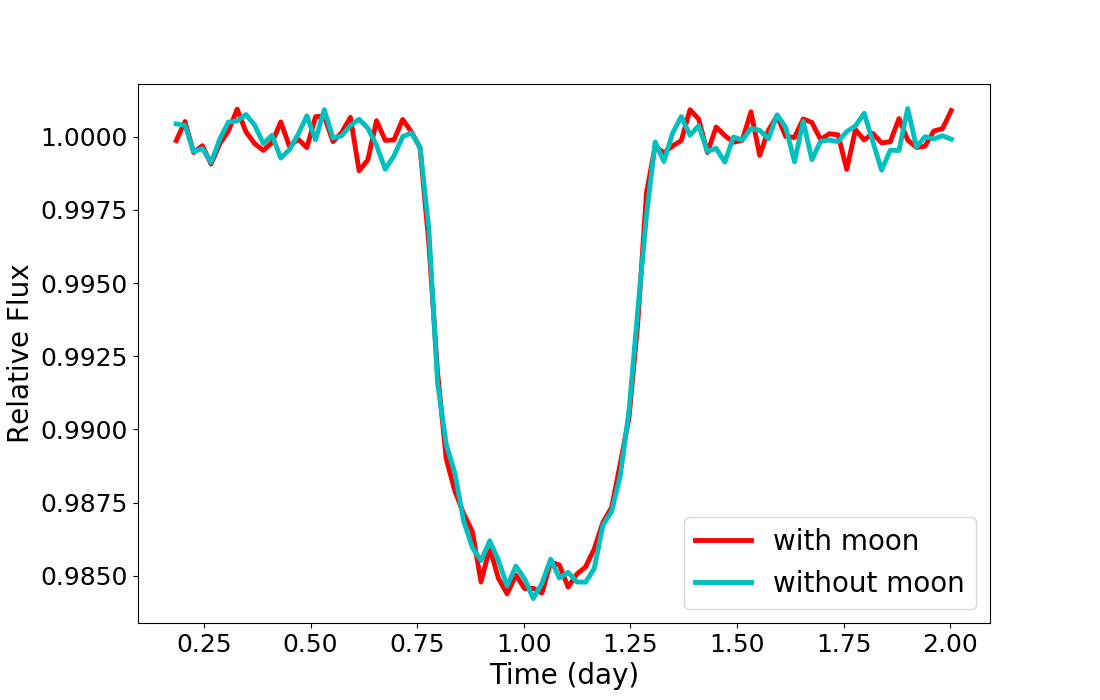}
\caption{Examples of simulated transit light curves. {\it Left panel:} The blue curve shows a transit of a planet with radius $R_{\rm p}=12 \text{R}_\oplus$ (similar to that of Kepler 1625b) for a low level of noise (100~ppm). The red curve shows a similar but different planetary transit, and an exomoon transit is superimposed ($R_{\rm m}=4.75\text{R}_{\oplus}$).
{\it Right panel:} Simulations for a higher noise level (400~ppm) and a much smaller moon $R_{\rm m}=\text{R}_{\oplus}$. The distinction between the planet-only (blue) and planet-moon (red) case is not visible by eye in most cases.}
\label{fig:lightcurve}
\end{figure*}

We propose a  solution based on a convolutional neural network (CNN) to detect exomoons in photometric light curves. To test the method, we generate synthetic transit light curves with and without an exomoon. The light curves are similar to those of the Kepler-1625~b observations. Fig.~\ref{fig:lightcurve} presents examples of such simulated light curves. The left panel shows a planetary transit including an exomoon at low noise level. The right panel shows a transit containing an exomoon at  higher noise level.  The exomoon transit is always close in space to the planet transit because the exomoon is on a close-in orbit of the planet. Depending on the orbital configuration, the exomoon transit can occur before or after the planetary transit.

Recently, deep CNN architectures have achieved impressive performance for data classification. They are quickly becoming prominent in astrophysical applications because they have the ability to effectively encode spectral and spatial information based on the input data, without pre-processing. A typical network consists of multiple interconnected layers and learns a hierarchical feature representation from raw data~\citep{vgg:2014}. The network optimization is usually performed with L2-norm without considering the effect of the noise properties, and it does not exploit neighboring information in the data \citep{JanochaC17}. We here propose a 1D CNN to classify a transit signal as either containing or excluding an exomoon candidate. The architecture consists of five 1D convolution layers, followed by one fully connected layer. Gaussian smoothness is used as total variation loss to penalize the L1-norm error. The goal of using this smoothness is to remove the noise or unwanted variations in the data (e.g., from stellar activity, or systematic or instrumental effects) and conserve the neighborhood information at the same time.

{Deep learning has led to significant advances in the field of astrophysics. Most of the studies that incorporate this method were based on well-known network configurations, such as alexnet~\citep{alex:2012} and vgg16~\citep{vgg:2014}, which is tuned using a mean standard error as cost function. We here propose an improved cost function that includes a total variation loss component to regularize the network. Total variation loss has been used extensively in other fields, for example, in computer vision. \cite{mahendran2015} used the total variation as a regularization procedure for reconstructing natural images from their representation based on image priors. \cite{mehran2016} used the total variation as a structured loss by applying a Sobel edge-detection technique on the output probability map to train a deep network, especially in the case when there are not enough labeled data. We apply similar ideas to improve the classification of exoplanet light curves.}

The remainder of this paper is organized as follows. In section~\ref{Dataset} we describe the data sets we used in the analysis. An overview of our proposed method is given in Section~\ref{sec:method}. Section~\ref{sec:result} compares the performance of the proposed method with the performances of several other techniques in both generalization and convergence. Section~\ref{sec:summary} summarizes the most important findings.

\section{Synthetic data}
\label{Dataset}
We used three sets of light curves (Table~\ref{tab:dataset2}). Each light curve contained four transits. For each transit we had 50 days of data, with 25 days on each side of a transit. The light curves had a cadence of 29.4 minutes, which is Kepler's long cadence. The simulated transit light curves were generated using the model described in detail by \cite{kai_paper}, with the additional improvement that the occultation of the moon by the planet was taken into account.

The first data set consists of 1,052,400 simulated light curves; 526,200 with exomoons and 526,200 without exomoons. The noise was varied between 25~ppm and 500~ppm. The radius was also varied between 0.25 R$_{\oplus}$ to 5 R$_{\oplus}$. The exomoon phase was randomized between 0 and 1 for each light curve that contained a moon.

The second data set consisted of ten subsets. Each subset had 200,000 curves: 100,000 with {exomoons} and 100,000 without {exomoons}. The noise was kept constant at 100~ppm. The exomoon radius was varied from 0.25 R$_{\oplus}$ to 4.75 R$_{\oplus}$ from one subset to another. The exomoon phases were also randomized.

The third data set also contained ten subsets, each with 200,000 curves. There was either no moon (100,000) or a moon (100,000) with a radius of 1 R$_{\oplus}$. The noise was varied from 25~ppm to 475~ppm. The exomoon phases were again randomized.
\begin{table}[t]
\caption{Parameters for the three data sets of simulated light curves. The second and third columns give the range of values for the {exomoon} radius and the noise level, respectively. The last column gives the total number of simulated light curves, $M$.}
\centering
\begin{tabular}{llll}
\hline\hline
Data set & Exomoon radius & Noise level & $M$ \\ \hline
1& $0.25$\,--\,$5 R_{\oplus}$ & 25\,--\,500~ppm & 1052400 \\
2 & $0.25$\,--\,$4.75 R_{\oplus}$ & 100~ppm & 10 $\times$ 200000 \\
3 & $1 R_{\oplus}$ & 25\,--\,475~ppm & 10 $\times$ 200000 \\ \hline
\end{tabular}
\label{tab:dataset2}
\end{table}

Table~\ref{tab:dataset1} summarizes the choice of astrophysical parameters for the simulations, that is, the  stellar, planetary, and exomoon parameters. The stellar parameters were fixed; the star had  the same radius $R_\star$ and mass $M_\star$ as our Sun. We used a quadratic limb-darkening law, which describes the dimming of a star's brightness from the center of the stellar disk toward the limb (see~\citep{2011A&A...529A..75C} for more detail), with roughly solar-like values. The exoplanet radius $R_{\rm p}$ and mass $M_{\rm p}$ are roughly Jupiter-like, and the orbital distance was $0.85$ astronomical units. The exoplanet orbits its star in  287~days, in accordance with Kepler's third law. This law describes the relation between the distance of a planet to its star, the orbital period of the planet, and the combined mass of the star, the planet, and the moon if present, to determine the distance from the star. The exomoon semimajor axis around its planet $a_{\rm m}$ was set to 20~R$_{\rm Jupiter}$. The planet mass and the exomoon semimajor axis determine the orbital period of the moon around its planet. We varied the exomoon radius between 0.25 and 5 R$_{\oplus}$ while keeping its mass density constant and equal to the Earth's density. We chose to keep the density constant, therefore only one exomoon parameter was varied. We also added noise of various amplitudes between 25~ppm (parts per million) and 500~ppm to the synthetic light curves, instead of using a more realistic noise model. The initial exomoon phase $\varphi_0=\varphi(t=0)$ (see Fig.~\ref{fig:orbital_configuration}) was randomized for each realization. The realization means a new generation of the light curve with the appropriate parameters, with newly generated noise added to the model. The $q_1$ and $q_2$ determine the brightness of the star as a function of center-to-limb distance. Limb darkening affects the shape of the transit: the bottom of the transit is rounded and not flat (see Fig.~\ref{fig:lightcurve}).
\begin{table}[t]
\caption{Stellar, planetary, and exomoon parameters for the transit light-curve simulations. The subscripts $\odot$ and $\oplus$ refer to the Sun and Earth values, respectively.}
\centering
\begin{tabular}{ll}
\hline\hline
Parameter & Value (or range)\\ \hline
Planet-star radius ratio, $r_{\rm p}$ & $0.11$\\
Barycenter semimajor axis,  $a_{\rm b}$ & $183 R_\star$ \\
Barycenter impact parameter $b$ & 0\\
Barycenter initial orbital phase $\varphi_{\rm b}$ & 0 \\
Barycenter orbital period $P_{\rm b}$ & 287\,days \\
Stellar limb-darkening parameters & $q_1=0.6$, $q_2=0.4$  \\
Moon-star radius ratio, $r_{\rm m}$ & $0.00229$ -- $0.04584$\\
Moon semimajor axis, $a_{\rm m}$ & $2.055 R_\star$ \\
Moon initial orbital phase, $\varphi_{\rm m}$ &  0\,--\,$2\pi$ (random)\\
Moon orbital period, $P_{\rm m}$ & $0.5523$ days\\
Moon-planet mass ratio, $f_{\rm m}$ & $1.2\times 10^{-5}$ -- $0.0942$ \\
\hline
\end{tabular}
\label{tab:dataset1}
\end{table}

\section{Method}
\label{sec:method}
In this section, the proposed framework for detection of a moon-like signal from a light curve is described. The method does not require a preprocessing step. The 1D CNN architecture is summarized below and described in full in  Appendix ~\ref{sec:Network_Architecture}. Our definition of the loss function is presented below. 

{The idea of detecting a moon in light curves is similar to the idea of detecting an edge in images, where brightness (pixel value) changes sharply. The convolution stage in CNNs involves the sharing of connection weights, and this allows for robust position-invariant feature detection. Time variations are detected  regardless of when they happen to be within the window of interest. }

The input of the CNN is a simulated light curve, and the output is a binary value indicating the presence of a moon. We denote the $m$th simulated light curve by $X_m = \{X_m(0), X_m(1) \cdots X_m(N-1) \}$, where $0\leq m \leq M-1$. Each light curve contains four transits. The total number of time samples is $N=9796,$ and the number of simulated light curves, $M$, is given in Table~\ref{tab:dataset2}. 
For each input simulated light curve, the output of the network is denoted by $C(X_m)$, equal to 0 (no moon) or 1 (moon). The true classification of the light curve is $c_m=1$ (moon) or $c_m=0$ (no moon). We note that convolutional networks are also appropriate because they are computationally efficient; the dimension of each input is reduced by convolution and pooling stages.

\subsection{Convolutional neural network architecture}
\label{sec:method1}
\begin{figure*}[t!]
\includegraphics[width=\textwidth,height=50mm]{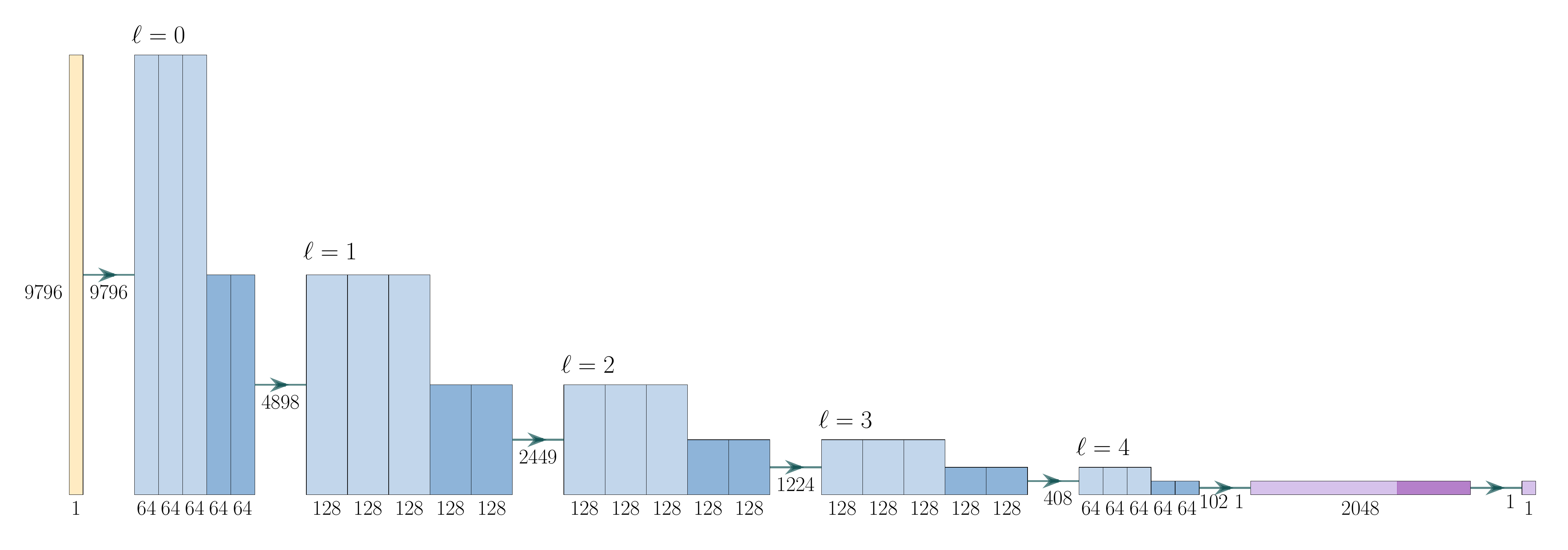}
\caption{The CNN architecture. As explained in the text, it consists of five convolution layers composed of 64, 128, 128, 128, and 64 channels, including convolution, batch normalization, leaky ReLU activation {(first three columns in light blue)}, maximum pooling, and Gaussian dropout {(last two columns in dark blue)}. The last layer is followed by a fully connected layer {(purple)} and a classification layer returning a binary answer (1 for moon, 0 otherwise).}
\label{fig:arch}
\end{figure*}

Our network takes $X_m$ as input and classifies it into a binary value. As presented in Figure~\ref{fig:arch}, the architecture of the network consists of five convolution layers, followed by one fully connected layer and one output layer. Each convolution layer consists of a 1D convolution, batch normalization, leaky ReLU activation, maximum pooling, and Gaussian dropout. {The determination of the network architecture was obtained experimentally.}

\subsection{Loss function}
\label{sec:method2}
The training process of the CNN is accomplished through the minimization of a loss function that measures the error between correct and predicted values. In the classification problems, there are many loss functions that could be used (e.g., binary cross-entropy, hinge, Kullback–Leibler (KL) divergence, and wasserstein)~\citep{JanochaC17}.

Noise-smoothing operators have been extensively used as preprocessing in the context of signal processing to suppress the noise and preserve the changes at the same time~\citep{smoothing1}. The contribution of our work is the use of the smoothing operator directly in learning. The loss function is constructed using spatial information (e.g., neighborhood information) of the data features. As a result, back-propagation errors of the proposed loss functions are different. In this section, we define the proposed loss function that combines mean error and total variation error.

We consider a binary classification, where the goal is the assignment of either no-moon (0) or moon (1) to each moon-like signal in the training set. The input data are a set of training samples indexed by $0\leq m \leq M-1$. The loss function is a combination of two functions. The first function is the mean error,
\begin{equation}
L_{\rm err}(\Theta)= \frac{1}{M}\sum_{m=0}^{M-1}
| C(X_m; \Theta) - c_m |,
\label{eq:mse}
\end{equation}
where each term in the sum can only take the values 1 or 0. This function is to be minimized to obtain all the parameters $\Theta$ of the CNN (all weights and biases). However, if the number of parameters $\Theta$ is large and the amount of data is low, the problem might be overfit. Regularization by penalizing the loss function is a common solution for this problem. 

The output of the last convolution layer ($\ell=4$) is a matrix of dimension 64 (number of feature maps) times 102 (number of time samples). This layer is followed by a fully connected layer $Y(p)$ activated by a leaky ReLU (see Appendix). We use the total variation of $Y$ as a penalty to regularize the classification. First, each vector $Y_m$ is smoothed to remove noise while preserving the original signal. This is done by convolution with a centered Gaussian kernel $G$ with standard deviation $\sigma=3$ and width $9$,
\begin{equation}
S_m(p) = \sum_{j=0}^{8} Y_m(p-j)G(j).
\end{equation}
The total variation loss is then obtained by computing the mean unsigned difference between two consecutive points, and averaging over all the light curves,
\begin{equation}
L_{\rm tv}(\Theta) = \frac{1}{M} \sum_{m=0}^{M-1}  \frac{1}{N_p-1}  \sum_{p=0}^{N_p-2} \left |(S_m(p+1) - S_m(p) \right|.
\end{equation}

The final loss function is a linear combination of the mean error between the correct and predicted values and the average unsigned gradient of the input features,
\begin{equation}
L(\Theta) =  L_{\rm err}(\Theta) +  \gamma L_{\rm tv}(\Theta),
\label{eq:loss}
\end{equation}
where $\gamma$ is a regularization parameter. We here give equal weight to the two terms, that is, $\gamma=1$.

\subsection{Optimization}
In order to obtain the network parameters $\Theta^*$ that minimize $L(\Theta)$, we use the AdamMax algorithm to gradually update the weights and the biases when searching for the optimal solution, see Appendix ~\ref{sec:Optimization}.

\section{Performance}
\label{sec:result}
We present evaluation metrics and setup (Sect.~\ref{Performance_Metrics}) to assess the performance of the CNN in comparison to the classical method for the detection of exomoons (Sect.~\ref{Classic_Analysis}). In section~\ref{Experimental_Analysis} the performance of the proposed method is further discussed by varying the parameters of the moon, the planet, and the host star.

\subsection{Performance metrics and setup}
\label{Performance_Metrics}
The most common metrics for the evaluation of a binary classification are sensitivity, specificity, and accuracy. Sensitivity, also called the true-positive rate, measures the proportion of actual positive samples that are correctly classified as positive samples. Specificity, also called true-negative rate, is the proportion of actual negative samples that are correctly classified as negative examples~\citep{Hand:2001}. Accuracy measures the proportion of correctly classified light curves. These metrics are defined as follows:
\begin{eqnarray}\label{eq:Completeness}
&\text{Sen} =& \frac{TP}{TP+FN}  , \qquad\qquad \quad \qquad\text{(sensitivity)} \\
\label{eq:Correction}
&\text{Spe}  = &\frac{TN}{TN+FP} ,  \qquad\qquad \quad  \qquad\text{(specificity)}  \\
\label{eq:Quality}
& \text{Acc} = &\frac{TP+TN}{TP+FP+FN+TN} ,  \qquad\!\!\!\! \text{(accuracy)}
\end{eqnarray}
where $TP$ is the number of true positives (exomoons correctly detected), $FP$ is the number of false positives, $TN$ is the number of true negatives (absence of moons correctly classified), and $FN$ is the number of false negatives.

We used 70$\%$ of each data set (in Table~\ref{tab:dataset2}) to train and validate the CNN and $30\%$ of data to evaluate the CNN performance using the above metrics.
All our CNN experiments were run on Nvidia Tesla V100 GPUs with Keras 2.2.2 with Tensorflow 1.9.1. Each CNN experiment with the same configuration was run 25 times. The results we present here were obtained by averaging over these 25 trials. The code of the network can be found at \url{https://github.com/ralshehhi/Exomoon}.

\subsection{Classical analysis}
\label{Classic_Analysis}
For comparison with the CNN algorithm, we also classified the light curves using the classical method described in~\citep{kai_paper}. For each light curve, we considered two models that might explain the data: a light-curve model containing the transits of a single planet without a moon (which we call no-moon model), and a model containing the transits of both a planet and a moon (which we call one-moon model). To estimate the likelihood of each case, we used the Bayesian information criterion \citep[BIC,][]{1978AnSta...6..461S} to determine whether a moon is present. The BIC of a model ${\cal M}(\theta)$, which depends on model parameters $\theta=\{\theta_1, \theta_2 \cdots \theta_k\}$, is given by
\begin{align}
    \text{BIC}({\cal M})=-2\max_\theta{\mathcal{L}(\theta|X)}+k\ln{N},
\end{align}
where $\mathcal{L}(\theta|X)$ is the likelihood of $\theta$ given the data $X$, $k$ is the number of used parameters ($k=14$ for the one-moon model and $k=7$ for the no-moon model), and $N=9796$ is the length of $X$, as described in Sect.~\ref{sec:method}. We calculated the difference $\Delta$BIC $=$ BIC (one moon) $-$ BIC (no moon) between the two BICs and classified light curves with a $\Delta{\rm BIC}>0$ as containing no moon and $\Delta{\rm BIC}<0$ as containing a moon. The $\Delta$BIC between two competing models compared the maximum likelihood of the two models while penalizing the model with more parameters. In other words, the model with more parameters has to explain the data sufficiently better to justify the use of more free parameters. The one-moon model contains five more parameters than the no-moon model, which describes the moon size and orbital configuration (e.g., ratio of the moon-to-star radius, planet-moon distance, moon period, moon phase, and ratio of moon-to-planet mass). We find the maximum likelihood of the model with a moon and without a moon using the Markov chain Monte Carlo (MCMC) sampler \textit{emcee}, which is used to approximate the parameter posterior distribution $\mathcal{L}(\theta|X)$.

\subsection{Results and analysis}
\label{Experimental_Analysis}
\subsubsection{Effect of the choice of loss function}
In order to illustrate the effectiveness of the total variation loss function $L$ defined in section~\ref{sec:method2}, we considered other loss functions: binary cross-entropy, hinge, KL, wasserstein, and mean square error functions.
Tables~\ref{tab:Comparison_LossFunctions1} and~\ref{tab:Comparison_LossFunctions2} show faster convergence and better performance of $L$ than all other choices of loss functions.
\begin{table}[t]
\caption{Number of epochs needed to train the model and computing time in second required to process the test data for different choices of loss functions. Each epoch consists of 32 samples from data set 1.}
\centering
\begin{tabular}[t]{lll}
\hline\hline
Loss function & Training &
Testing \\
& (number of epochs) &
(Computation time (s))\\ \hline
{Binary entropy} & 24   & 149.7 \\
{Hinge}          & 100  & 148.9 \\
{KL}             & 100  &  148.3 \\
{Wasserstein}    & 100  & 145.7 \\
$L_{\rm err}$            & 20   & 146.8\\
($L_{\rm err}+L_{\rm tv}$)  & \textbf{23} & \textbf{146.8} \\ \hline
\end{tabular}
\label{tab:Comparison_LossFunctions1}
\end{table}
\begin{table}[t]
        \caption{Comparison between loss functions based on performance metrics: sensitivity, specificity, and accuracy. Data set 1 was used.}
        \centering
        \begin{tabular}{llll}
        \hline\hline
        Loss function& Sen& Spe& Acc \\
        &($\%$)&($\%$)&($\%$) \\
        \hline
        {Binary entropy}      & 97   &  49  & 73 \\
        {Hinge}       & 100    &  0  & 50 \\  %
        {KL}          & 100  &    0 & 50 \\  %
        {Wasserstein} & 0  &   100  & 50 \\  %
        {$L_{\rm err}$}         &  90  &  96 & 93 \\  %
        {($L_{\rm err}$+$L_{\rm tv}$)}  &  \textbf{94}  & \textbf{97} &  \textbf{96} \\ \hline
        \end{tabular}
        \label{tab:Comparison_LossFunctions2}
\end{table}

\subsubsection{Effect of smoothing as preprocessing}
\label{sec:pre-whitening}
The main purpose of the preprocessing step is to produce more effective features by standardizing the dynamic range of the raw data or to remove unwanted variations in the raw data. Smoothing is one of the preprocessing steps that might be used to remove unwanted variations. However, this approach could introduce new systemic variability or remove actual signal from the raw data. In Table~\ref{tab:Comparison_prewhitening} we compare the smoothing method as a preprocessing step of 1D CNN with $L_{err}$ and the method we propose here, which uses $L_{tv}$ to learn the parameters.
\begin{table}[t]
\caption{Comparison between smoothing techniques. The first four rows show performance results obtained by applying smoothing (with different kernels) before the CNN using $L_{\rm err}$ to train the network. The last row presents the performance metrics of CNN with the loss function (eq.~\ref{eq:loss}) without smoothing. Data set 1 was used.}
\centering
\begin{tabular}{llll}
\hline\hline
Method  & Sen &Spe &Acc \\
&  ($\%$)& ($\%$)& ($\%$) \\
\hline
{Median + $L_{\rm err}$}  &   81  &  99  & 90 \\
{Cosine + $L_{\rm err}$}  &   92  &  90  & 91 \\
{Hamming + $L_{\rm err}$}  &  93   &  81  &  87\\
{Gaussian + $L_{\rm err}$}  &   92  &  88  & 90  \\
\textbf{No prefilter + $L_{\rm err}$ + $L_{\rm tv}$}  & \textbf{94}  & \textbf{97} &  \textbf{96} \\ \hline
\end{tabular}
\label{tab:Comparison_prewhitening}
\end{table}

\subsubsection{Dependence on exomoon radius}
\label{sec:radius}
\begin{figure}[t]
\includegraphics[width=1.11\columnwidth]{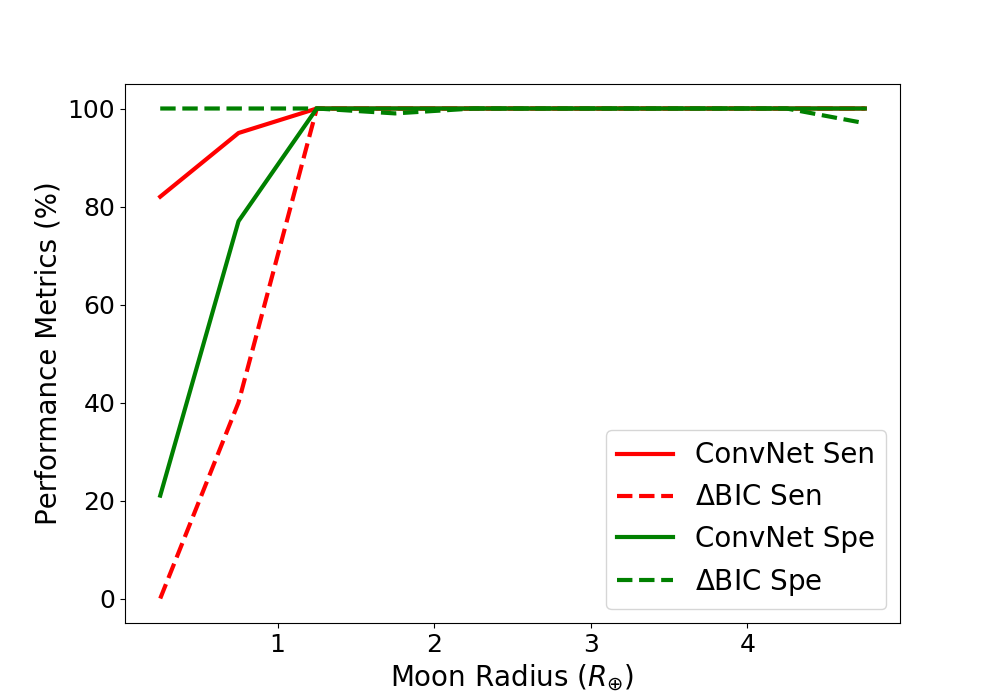}
\caption{Performance as a function of exomoon radius at fixed noise level (100~ppm). Data set 2 was used.}
\label{fig:Radius_PerformanceMetrics}
\end{figure}

Figure~\ref{fig:Radius_PerformanceMetrics} shows the results we obtained by increasing the exomoon radius. Sen and Spe are significantly lower when the exomoon radius is smaller than the Earth radius, although the convolution, pooling, and smoothing windows are small enough to capture small differences in data distribution. When the exomoon radius is equal to or greater than the Earth radius, the performance reaches  100$\%$. The total variation is commonly used as a regularization factor to find continuity in the  signals~\citep{houhou2009} and detect a change between neighboring values, but the variations due to small moons are very difficult to identify.

The addition of a small moon alone alters the light curve slightly, which also means that the maximum likelihoods of the no-moon and one-moon models are very similar. In this case, the $\Delta$BIC method decides in favor of the model with the fewer parameters, and Sen goes to zero.

\subsection{Effect of noise on the light curves}
\label{sec:noise}
\begin{figure}[t]
\includegraphics[width=1.11\columnwidth]{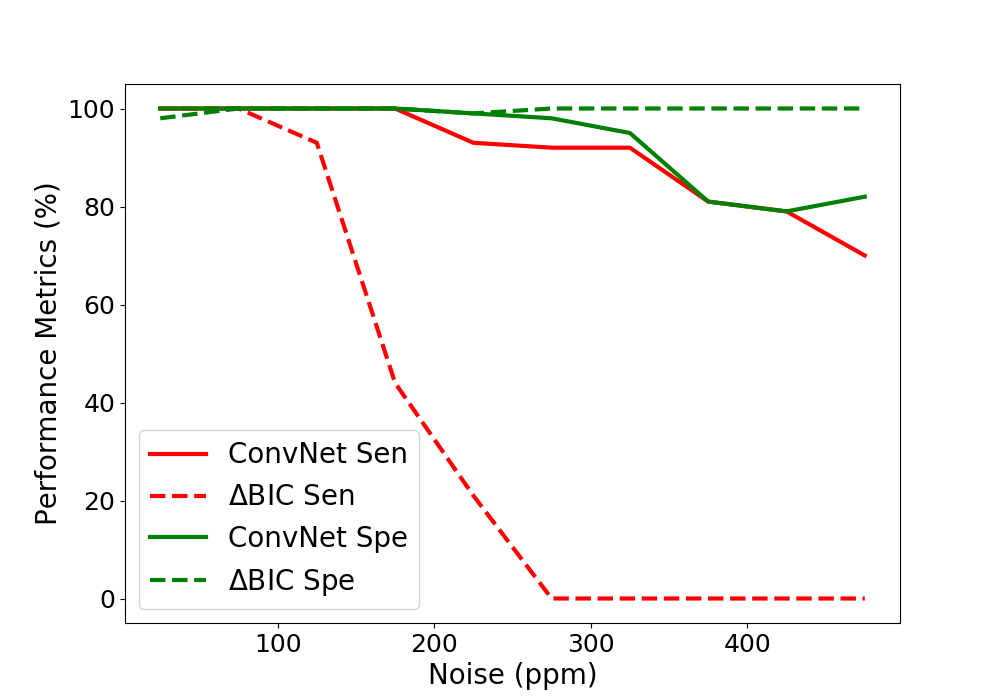}
\caption{Performance as a function of noise level between 25~ppm and 475~ppm. The exomoon radius is $1 R_\oplus$. Data set 3 was used.}
\label{fig:Noise_PerformanceMetrics}
\end{figure}

Figure~\ref{fig:Noise_PerformanceMetrics} shows the results for detecting a $1 R_\oplus$ moon when the amount of noise is varied from 25~ppm to 475~ppm every 50~ppm. We find that when the noise is lower than 175~ppm, the performance of the network is perfect, and it is very good (above 80 \%) when the noise is below 375~ppm. The performance decreases to Acc$=75\%$ for the highest noise level. The sensitivity is higher for the CNN than $\Delta$BIC for two reasons. First, the total variation~\citep{tv_} enforces spatial constraints, which helps the learning process to define spatial continuity in the signal and distinguish between a moon-like signal and noise. Second, adding Gaussian noise with dropout also helps to detect changes in the signal (see, e.g.,~\citep{dropout}).

On the other hand, for high noise levels, the difference in the maximum likelihood between the one-moon model and the no-moon model is too small to counteract the penalty given by the difference in the number of parameters between the two models.

\subsection{Effect of training size}
\label{sec:number_samples}
\begin{figure}
\includegraphics[width=1.11\columnwidth]{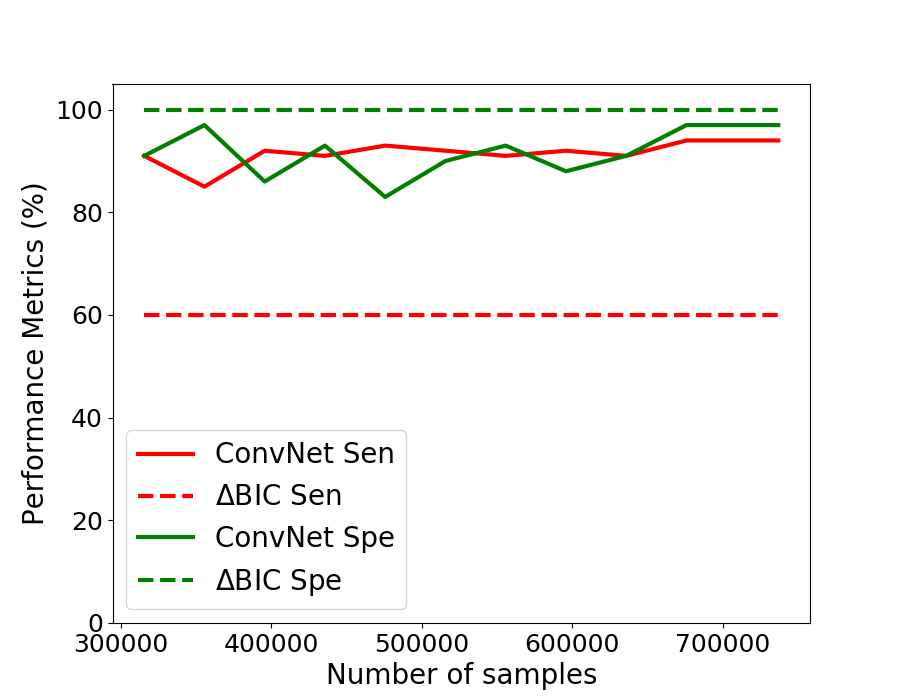}
\caption{Results as a function of training size. The training set includes all noise levels and exomoon sizes. Data set 1 was used.}
\label{fig:Size_PerformanceMetrics}
\end{figure}

Figure~\ref{fig:Size_PerformanceMetrics} presents the performance metrics as functions of the number of samples in the training set from 300,000 to 700,000.
The number of samples for the testing set was kept constant at 315,720 samples. As expected, the CNNs are more robust when larger training sets were used. The results show that training sets with 650,000 samples achieve only little improvement. For the largest three training sets shown in Fig.~\ref{fig:Size_PerformanceMetrics}, the CNN performance appears to have converged.

\subsection{Comparison with the $\Delta$BIC method and other CNN architectures}
\label{sec:comparison}
Table~\ref{tab:Comparison_networkArchitectures} shows the comparison between the method we propose here, the classic approach for detecting moon-like signals ($\Delta BIC$), and other well-known CNN architectures (Vgg16 1D~\citep{vgg:2014}, AlexNet 1D~\citep{alex:2012}, ResNet 1D~\citep{resnet}, and DenseNet 1D~\citep{dense}). The difference between the approach we propose and the other CNN approaches is in the size of the kernel filters, the number of filter units, the number of convolution layers, and the order of convolution layers. The results of our proposed CNN with total variation loss outperforms all other approaches. The classical approach ($\Delta$BIC) performs more poorly than all CNNs.

The $\Delta$BIC method requires around 12\--\,18 hours per light curve on one CPU with a memory of 32 GB per core, which requires above $2\times10^{10}$ s to analyze 315720 light curves (test data). On the other hand, the CNN requires less than one second per light curve on one GPU with 32 GB per core, which corresponds to only about $200$ seconds for the entire test data set.
\begin{table}[t]
\caption{Performance comparison between the proposed CNN and other 1D CNN architectures. The last column gives the computing time required to process the test data. Data set 1 was used.}
\centering
\begin{tabular}{lllll}
\hline\hline
Method & Sen & Spe & Acc & Computation time\\
&($\%$)& ($\%$)& ($\%$) &  (s)\\
\hline
{$\Delta$BIC}      &  60 & 100 & 80  &  $>2\times 10^{10}$ \\
{Vgg16}      & 97    &  86  &  86 & 164 \\
{AlexNet}  & 0     &  100  & 50 & 90 \\
{ResNet}  & 44    &  100  & 75 & 170 \\
{DenseNet} &  44   &  100  & 75 & 180 \\
\textbf{Proposed}    & \textbf{94}  & \textbf{97} &  \textbf{96} & \textbf{158} \\ \hline
\end{tabular}
\label{tab:Comparison_networkArchitectures}
\end{table}

\subsection{Training the CNN using a more general data set}
The synthetic data from the previous sections were all generated using only three free parameters (exomoon radius, noise level, and exomoon phase), while the other parameters of the star and planet were kept constant. In order to test how the CNN method behaves when it is trained on a more general data set, we generated new synthetic light curves for which we varied all 14 free parameters (Table~\ref{tab:dataset4}). This new data set includes 200000 light curves. 
\begin{table}[t]
\caption{Stellar, planetary, and exomoon parameters for the transit light-curve simulations with 14 free parameters.
}
\centering
\begin{tabular}{ll}
\hline\hline
Parameter & range \\ \hline
Planet-star radius ratio, $r_{\rm p}$ & $0.065$ -- $0.126$\\
Barycenter semimajor axis, $a_{\rm b}$ & $49.5 R_\star$ -- $281.7 R_\star$\\
Barycenter impact parameter, $b$ & $-1$ to $1$\\
Barycenter initial orbital phase, $\varphi_{\rm b}$ & 0\\
Barycenter orbital period, $P_{\rm b}$ & $200$ -- 300\,days \\
Stellar limb-darkening parameters & $q_1=0.4$ -- $0.6$\\
 &$q_2=0.4$ -- $0.6$  \\
Moon-star radius ratio, $r_{\rm m}$ & $0.0065$ -- $0.0504$\\
Moon semimajor axis, $a_{\rm m}$ & $10^{-5}R_\star$ -- $0.005R_\star$\\
Moon initial orbital phase, $\varphi_{\rm m}$ & full range \\
Moon orbital period, $P_{\rm m}$ & $0.5$ -- 10 days\\
Moon-planet mass ratio, $f_{\rm m}$ & $0.001$ -- $0.2$\\
Moon orbital inclination & full range \\
Moon argument of ascending node & full range \\
Noise level & 50 -- 500 ppm \\ \hline
\end{tabular}
\label{tab:dataset4}
\end{table}

\begin{figure}[t]
\includegraphics[width=1\columnwidth]{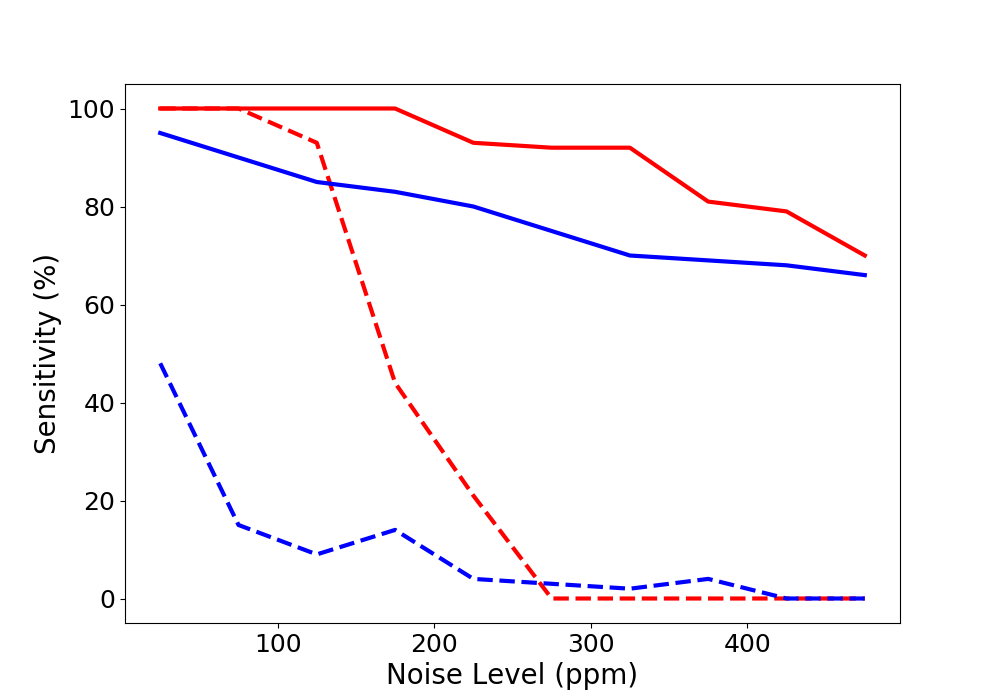}
\caption{Sensitivity as a function of noise level using input data generated with 3 free parameters (exomoon radii, exomoon orbital phases, and noise levels, see Table~\ref{tab:dataset1}; red curves) and with 14 free parameters (see Table~\ref{tab:dataset4}; blue curves).  The two methods are compared at a fixed exomoon radius of 1 $R_{\oplus}$: ConvNet (solid curves) vs. $\Delta$BIC (dashed curves).
The sensitivity is plotted in bins of 50 ppm noise levels.}
\label{fig:noise_sen_3_14}
\end{figure}

Figure~\ref{fig:noise_sen_3_14} shows the sensitively of both $\Delta$BIC and the CNN versus noise level at a fixed exomoon radius of $1 R_\oplus$. 
When the CNN is trained using the dataset with 14 free parameters, the performance of the CNN decreases by up to $20\%$ compared to its performance when the more restricted training data set is used. It is important to note that the CNN still performs well at high noise levels.
For noise levels above about 200 ppm, the performance of the $\Delta$BIC method is far worse than the CNN, independently of the data set it is applied to.

\section{Conclusion}
\label{sec:summary}
We proposed a regularized 1D CNN architecture to detect exomoon candidates in transit light curves. The regularized loss function brings a significant improvement over traditional loss functions in terms of convergence, sensitivity, and specificity. Furthermore, the method we propose performs relatively well with high levels of noise and small moon sizes.

We compared the network results with the $\Delta$BIC method described by \cite{kai_paper}. The $\Delta$BIC method is more conservative: it has a higher true-negative rate, but also a much lower true-positive rate. For low noise levels (100 ppm), both methods are able to reliably detect exomoons larger than 1~R$_\oplus$  in our simulation setup. The CNN shows its strength at higher noise levels. For the data set with only 3 free parameters, Sen and Spe stay above $90\%$ up to a noise level of 325~ppm and remain around 80\% for even higher noise levels. For the data set with 14 free parameters, Sen remains above 65\% at 500 ppm. The $\Delta$BIC method is completely insensitive to the presence of exomoon at noise levels of 275~ppm and higher. The CNN is more effective in predicting a moon-like signal, and it is faster. It is a promising technique for applications to real astronomical data sets.
 
\begin{acknowledgements}
\newline
We thank Chris Hanson for constructive comments. This work was supported in part by NYUAD Institute grant G1502 ``Center for Space Science''. RA acknowledges support from the NYUAD Kawader Research Program. Computational resources were provided by the NYUAD Institute through Muataz Al Barwani and the HPC Center. The simulated light curves were produced at the DLR-supported PLATO Data Center at the MPI for Solar System Research.  {Author Contributions:} LG proposed the research idea. RA proposed the Machine Learning algorithm and analyzed the data and the results. KR prepared the simulated data and performed the BIC analysis. KR is a member of the International Max Planck Research School for Solar System Science at the University of G\"ottingen. All authors contributed to the final manuscript.
\end{acknowledgements}

\bibliographystyle{aa}
\bibliography{aa} 

\newpage

\appendix

\section{Network architecture}
\label{sec:Network_Architecture}
The network consists of the following: convolution blocks (convolution, batch normalization, maximum pooling, leaky ReLU activation and Gaussian dropout), fully-connected block and classification block. 

\textit{First convolution layer ($\ell=0$):}
The input of the first convolution layer, $\ell=0$, is a vector $X_m$ of dimension $N$. The index $m$ of the light curve is fixed throughout this section, therefore we dropped it to simplify the notation. To construct feature maps, the input data were convolved $64$ times with different kernels $F_{k,0}$ of width $n^0_f=3$ and stride $s_f=1$. The kernel values are to be  determined later by training the network. The stride  is the distance used to shift the convolution kernel. We obtained $64$ feature maps of length $N_f^0=N$ (with zero padding), see, for example, \cite{alex:2012},
\begin{equation}
Y^{k,0}_{\rm conv}(i) =  \sum_{j=0}^{n^0_f-1} X(i-j) F_{k, 0}(j), \quad  0\leq k < K_0=64 .
\label{eq:conv}
\end{equation}

\textit{Batch normalization:}
The convolution operator is followed by a batch normalization function. Batch normalization is applied to reduce internal covariance shift~\citep{BN}. It is simply a linear transformation of the data with scaling $\lambda_0$ and a shift $\beta_0$,
\begin{equation}
Y_{\rm norm}^{k,0}(i) = \lambda_0 \frac{Y^{k,0}_{\rm conv}(i)-\mu_0}{\sqrt{v_0+\epsilon}} + \beta_0, 
\end{equation}
where $\mu_0$ and $v_{0}$ are the average and variance of the input data for all $k$s at fixed $\ell$. The initial values of the parameters were $\lambda_0=1$ and $\beta_0=0$. The constant $\epsilon=0.002$ was added to avoid division by very low values.

\textit{Leaky ReLU Activation:}
The batch normalization is followed by applying a leaky rectified linear unit (ReLU) activation function as follows~\citep{LeaklyReLU}:
\begin{equation}
\begin{aligned}
Y_{\rm lactiv}^{k,0} (i)
&=\text{LReLU}\left( Y_{\rm norm}^{k,0}(i) ; w_k, b_k \right)
\\
&= \begin{cases}
  Y_{\rm norm}^{k,0}(i) w_k+b_k & \text{ if } Y^{k,0}_{\rm norm}(i) \geqslant 0, \\
  {\rho} {Y}^{k,0}_{\rm norm}(i) w_k+b_k & \text{ else} .
\end{cases}
\label{eq:activation}
\end{aligned}
\end{equation}
Here $w$ and $b$ are values to be determined.
Many activation functions are possible, such as tanh, sigmoid, hyperbolic, or rectified functions. The leakly ReLU function allows for a small gradient {$\rho$} when the units are not active. Here, we set {$\rho=0.02$}, and refer to~\citep{LeaklyReLU} and~\citep{relu} for an in-depth discussion.

\textit{Maximum pooling:}
A maximum-pooling operator performs a spatial subsampling in time by taking the maximum value over a pooling window of length $n_p$ every $s_p$ points (with $n_p$ = $s_p$),
\begin{equation} Y^{k,0}_{\rm mpool}(i) = \max \{ Y_{\rm lactiv}^{k,0}(j) \}_{  i s_p \leq j \leq  i s_p +n_p -1}   ,
\label{eq:pmax}
\end{equation}
for $i=0,1, ...,    N_f^0/n_p-1$.

\textit{Gaussian dropout:}
In order to provide some regularization, a Gaussian dropout procedure is applied after the maximum pooling \citep{dropout}. It involves multiplying weights by a Gaussian random variable with  mean 1 and standard deviation $\sigma=0.65$.
Symbolically, we write
\begin{equation}
    Y^{k,\ell}_{\rm dout} = \text{\rm Gaussian$_{-}$dropout}
    (\ Y_{\rm mpool}^{k,\ell}\ ) .
\end{equation}
We found through experimentation that a Gaussian drop\-out gives better results than a standard dropout. 

\textit{Next convolution layers:}
Another four convolution layers ($\ell=1,2,3,\text{and }4$) follow the first convolution layer ($\ell=0$).  For each layer $\ell\ge 1$, we computed the feature maps,
\begin{equation}
Y_{\rm conv}^{k,\ell}(i) =  \sum_{j=0}^{n_f-1} Y_{{\rm  dout}}^{k,\ell-1}(i-j)  F^{k,\ell}(j), \quad 0\leq k < K_\ell,
\label{eq:convol}
\end{equation}
followed by batch normalization, leaky ReLU activation, maximum pooling, and dropout, as defined above.
The number of feature maps $K_\ell$ is 128 for the intermediate convolution layers $\ell=1,2,3$ and $K_4 =64$ for the last layer.

\textit{Fully connected layer:}
The output of the last convolution layer $\ell=4$ is a matrix of dimension 64 (number of feature maps) times 102 (number of time samples). This layer is followed by a fully connected layer $Y(p)$ activated by a leaky ReLU (see Eq.~\ref{eq:activation}),
\begin{equation}
    Y(p) = \sum_{k=0}^{63} \sum_{i=0}^{101} \text{LReLU}\left(  Y_{{\rm dout}}^{k,4}(i) ;  w_{k,i}(p), b(p)  \right),
\end{equation}
for $p\leq 0 \leq N_p-1,$ where $N_p = 64\times102=6528$ is the number of neurons; this is linearly converted to 2048.  The fully connected layer is the final vector output before the data are classified.

\textit{Output Layer: sigmoid activation and classification:}
This layer is activated by a sigmoid function, which produces a probability in the range between 0 and 1, as follows:
\begin{equation}
    \hat{y}  = (1+e^{-y})^{-1},
\end{equation}
with
\begin{equation}
y = \sum_{p=0}^{N_p-1} w_p Y(p) + b .
\end{equation}
The final classification of the data is
\begin{equation}
    C(X) =  \left\{
    \begin{array}{ll}
    1 & \text{ if } \hat{y} \ge  0.5, \\
    0 & \text{ else} .
    \end{array}
\right.
\end{equation}

\section{Optimization}
\label{sec:Optimization}
We used a batch approach, whereby the input samples are passed forward and backward through the network in chunks of 32 samples (in each epoch). We refer to \cite{adam} for additional details about this algorithm (AdaMax). Denoting $\Theta_{q-1}$ the parameters of the network at iteration $q-1$, the updated values at iteration $q$ are given by
\begin{equation}
\Theta_{q} = \Theta_{q-1} - \left(\frac{\alpha}{1-\beta_{1}}\right) \frac{m_q}{\mu_q},
\end{equation}
with
\begin{equation}
m_{q} = \beta_{1} m_{q-1} + (1-\beta_{1})\ \nabla L(\Theta_{q-1}),
\end{equation}
and
\begin{equation}
\mu_{q} = \max(\beta_{2}\mu_{q-1},\| \nabla L(\Theta_{q-1})\|) .
\end{equation}
The quantity $\alpha$ is the step size. $\eta = \alpha/(1-\beta_{1})$ is the learning rate, and the parameters $\beta_{1}=0.9$ and $\beta_{2}=0.999$ were fixed.

Initially, we set $\alpha=0.002$. At $q=0,$ we initialized the weights of the network in each layer with a random number drawn from a zero-mean Gaussian distribution with standard deviation $0.1$. The biases were initially set to zero.

A stopping criterion was used to reduce overfitting of the network.
Training stopped when the validation error did not decrease after 20 epochs.
If the loss function did not decrease, the parameter $\alpha$ was multiplied  by a factor of $0.1$ after the first 10 epochs, and the operation was repeated every 10 epochs.

\begin{figure}[t]
\includegraphics[width=1.11\columnwidth]{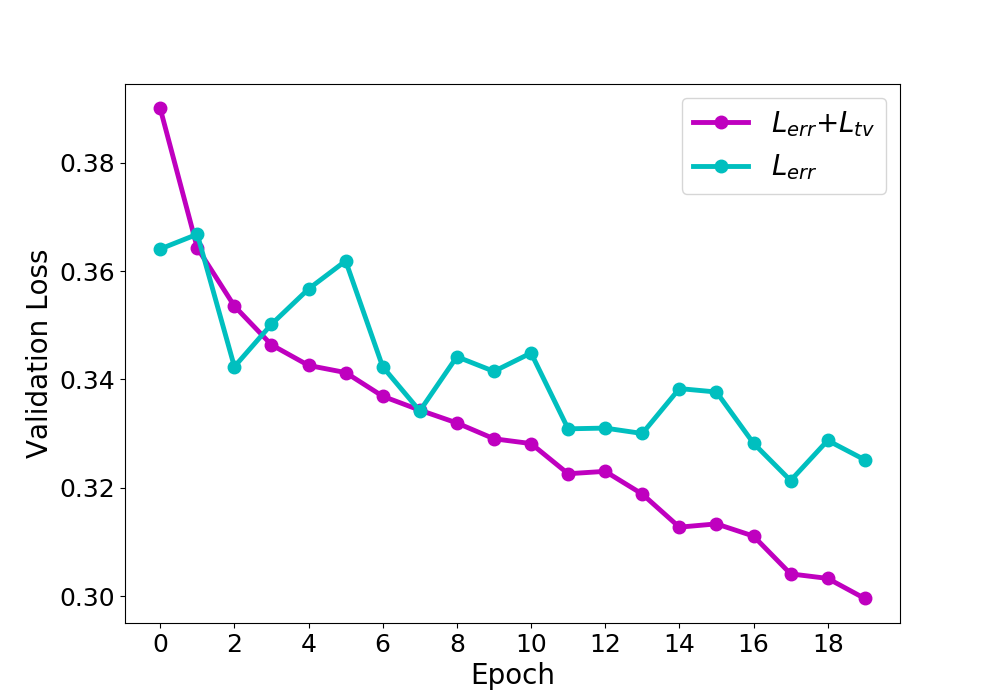}
\caption{Validation loss as a function of the number of epochs used to train the network. An epoch consists of 32 light curves. The regularized loss function $L_{\rm err} + L_{\rm tv}$ decreases faster than $L_{\rm err}$ alone.}
\label{fig:epoch_error}
\end{figure}

Figure~\ref{fig:epoch_error} compares the evolution of the loss function versus the number of epochs used to train the network. The regularized loss function performs the best.

\end{document}